\input harvmac
\overfullrule=0pt
%
\def\simge{\mathrel{%
   \rlap{\raise 0.511ex \hbox{$>$}}{\lower 0.511ex \hbox{$\sim$}}}}
\def\simle{\mathrel{
   \rlap{\raise 0.511ex \hbox{$<$}}{\lower 0.511ex \hbox{$\sim$}}}}
 
\def\slashchar#1{\setbox0=\hbox{$#1$}           
   \dimen0=\wd0                                 
   \setbox1=\hbox{/} \dimen1=\wd1               
   \ifdim\dimen0>\dimen1                        
      \rlap{\hbox to \dimen0{\hfil/\hfil}}      
      #1                                        
   \else                                        
      \rlap{\hbox to \dimen1{\hfil$#1$\hfil}}   
      /                                         
   \fi}                                         %
\def\ts{\thinspace}
\def\tx{\textstyle}
\def\ra{\rightarrow}

\def\ol{\bar}

\def\CA{{\cal A}}
\def\CB{{\cal B}}
\def\CC{{\cal C}}

\def\CF{{\cal F}}

\def\CL{{\cal L}}
\def\CM{{\cal M}}

\def\CO{{\cal O}}

\def\shat{\hat s}
\def\that{\hat t}
\def\uhat{\hat u}

\def\aqcd{\alpha_{S}}
\def\atro{\alpha_{\rho_T}}

\def\Ntc{N_{TC}}

\def\thw{\theta_W}
\def\kslash{\raise.15ex\hbox{/}\kern-.57em k}
\def\tro{\rho_{T1}}

\def\troct{\rho_{T8}}

\def\tpi{\pi_T}

\def\tpip{\pi_T^+}

\def\tpiz{\pi_T^0}

\def\octpi{\pi_{T8}}

\def\toppi{\pi_t}
\def\toppip{\pi_t^+}
\def\toppim{\pi_t^-}
\def\toppipm{\pi_t^\pm}
\def\toppiz{\pi_t^0}

\def\Mv{M_{V_8}}
\def\Mzp{M_{Z'}}
\def\tpilq{\pi_{L \ol Q}}

\def\tpiql{\pi_{Q \ol L}}

\def\jet{\rm jet}

\def\pbarp{\ol p p}

\def\gev{{\rm GeV}}
\def\tev{{\rm TeV}}

\def\pb{{\rm pb}}

\def\fb{{\rm fb}}
\def\half{{\textstyle{ { 1\over { 2 } }}}}

\def\twothirds{{\textstyle{ { 2\over { 3 } }}}}

\def\myfoot#1#2{{\baselineskip=14.4pt plus 0.3pt\footnote{#1}{#2}}}

\Title{\vbox{\baselineskip12pt\hbox{FERMILAB-CONF-96/298-T}
\hbox{BUHEP-96-34}
\hbox{hep-ph/9609298}}}
{Electroweak and Flavor Dynamics at Hadron Colliders--II}

\smallskip
\centerline{Estia Eichten\myfoot{$^{\dag }$}{eichten@fnal.gov}}
\smallskip\centerline{Fermi National Accelerator Laboratory}
\centerline{P.O.~Box 500 Batavia, IL 60510}
\centerline{and}
\smallskip
\centerline{Kenneth Lane\myfoot{$^{\ddag }$}{lane@buphyc.bu.edu}}
\smallskip\centerline{Department of Physics, Boston University}
\centerline{590 Commonwealth Avenue, Boston, MA 02215}
\vskip .3in

\centerline{\bf Abstract}

This is the second of two reports cataloging the principal signatures of
electroweak and flavor dynamics at $\pbarp$ and $pp$ colliders. Here, we
complete our overview of technicolor with a discussion of signatures
specific to topcolor-assisted technicolor. We also review signatures of
flavor dynamics associated with quark and lepton substructure. These occur
in excess production rates for dijets and dileptons with high $E_T$ and
high invariant mass. An important feature of these processes is that they
exhibit fairly central angular and rapidity distributions. This report
will appear in the Proceedings of the 1996 DPF/DPB Summer Study on New
Directions for High Energy Physics (Snowmass 96).

\bigskip

\Date{9/96}

\vfil\eject

\newsec{Introduction}

This and the preceding report summarize the major signals for dynamical
electroweak and flavor symmetry breaking in experiments at the Tevatron
Collider and the Large Hadron Collider. In the preceding report (referred
to below as~I), we reviewed the technicolor and extended technicolor
scenarios of dynamical electroweak and flavor symmetry breaking. We also
discussed signals for color-singlet and nonsinglet technipions, resonantly
produced via technirho and techni-omega vector mesons. In this report, we
complete this discussion with a summary of the main signatures of
tocolor-assisted technicolor: top-pions $\pi_t$ and the color-octet $V_8$
and singlet $Z'$ of broken topcolor gauge symmetries. These are presented
in section~2. This arbitrary division has been necessitated by the length
requirements of submissions to the Snowmass~'96 proceedings. In section~3
we motivate and discuss the main ``low-energy'' signatures of quark and
lepton substructure: excess production of high-$E_T$ jets and high
invariant mass dileptons. Cross sections are presented for a simple form of
the contact interaction induced by substructure. We re-emphasize that the
shapes of angular distributions are an important test for new physics as
the origin of such excesses. We also stress the need to study the effect of
other forms for the contact interactions. At the end of this report, we
have provided a table which summarizes for the Tevatron and LHC the main
processes and sample cross sections that we discussed in these two reports.

These reports are not intended to constitute a complete survey of
electroweak and flavor dynamics signatures accessible at hadron colliders.
We have limited our discussion to processes with the largest production
cross sections and most promising signal-to-background ratios. Even for the
processes we list, we have not provided detailed cross sections for signals
and backgrounds. Signal rates depend on masses and model parameters; they
and their backgrounds also depend strongly on detector capabilities.
Experimenters in the detector collaborations will have to carry out these
studies.

\newsec{Signatures of Topcolor-Assisted Technicolor}

The development of topcolor-assisted technicolor is still at an early stage
and, so, its phenomenology is not fully formed. Nevertheless, in addition
to the color-singlet and nonsinglet technihadrons already discussed, there
are three TC2 signatures that are likely to be present in any surviving
model; see Refs.~\ref\topcondref{Y.~Nambu, in {\it New Theories in
Physics}, Proceedings of the XI International Symposium on Elementary
Particle Physics, Kazimierz, Poland, 1988, edited by Z.~Adjuk, S.~Pokorski
and A.~Trautmann (World Scientific, Singapore, 1989); Enrico Fermi
Institute Report EFI~89-08 (unpublished)\semi V.~A.~Miransky, M.~Tanabashi
and K.~Yamawaki, Phys.~Lett.~{\bf 221B}, 177 (1989); Mod.~Phys.~Lett.~{\bf
A4}, 1043 (1989)\semi W.~A.~Bardeen, C.~T.~Hill and M.~Lindner,
Phys.~Rev.~{\bf D41}, 1647 (1990).},
\ref\topcref{C.~T. Hill, Phys.~Lett.~{\bf 266B}, 419 (1991) \semi
S.~P.~Martin, Phys.~Rev.~{\bf D45}, 4283 (1992);
{\it ibid}~{\bf D46}, 2197 (1992); Nucl.~Phys.~{\bf B398}, 359 (1993);
M.~Lindner and D.~Ross, Nucl.~Phys.~{\bf  B370}, 30 (1992)\semi
R.~B\"{o}nisch, Phys.~Lett.~{\bf 268B}, 394 (1991)\semi
C.~T.~Hill, D.~Kennedy, T.~Onogi, H.~L.~Yu, Phys.~Rev.~{\bf D47}, 2940 
(1993).},
\ref\tctwohill{C.~T.~Hill, Phys.~Lett.~{\bf 345B}, 483 (1995).},
\ref\tctwoklee{K.~Lane and E.~Eichten, Phys.~Lett.~{\bf B352}, 382
(1995) \semi
K.~Lane, Phys.~Rev.~{\bf D54}, 2204 (1996).},
\ref\hp{C.~T.~Hill and S.~Parke, Phys.~Rev.~{\bf D49}, 4454 (1994)\semi
Also see K.~Lane, Phys.~Rev.~{\bf D52}, 1546 (1995)\semi We thank
D.~Kominis for corrections to a numerical errors in both papers.}:

\item{$\circ$} The isotriplet of color-singlet ``top-pions'' $\toppi$ arising
from spontaneous breakdown of the top quark's $SU(2)\otimes U(1)$ chiral
symmetry;

\item{$\circ$} The color-octet of vector bosons $V_8$, called ``colorons'',
associated with breakdown of the top quark's strong $SU(3)$ interaction to
ordinary color;

\item{$\circ$} The $Z'$ vector boson associated with breakdown of the top
quark's strong $U(1)$ interaction to ordinary weak hypercharge.

\medskip

The three top-pions are nearly degenerate. They couple to the top quark
with strength $m_t/F_t$, where $m_t$ is the part of the top-quark mass
induced by topcolor---expected to be within a few GeV of its total
mass---and $F_t \simeq 70\,\gev$~\tctwohill\ is the $\toppi$ decay
constant.\foot{As far as we know, the rest of the discussion in this and
the next paragraph has not appeared in print before. It certainly deserves
more thought than has gone into it here. One possible starting place is the
paper by Hill, Kennedy, Onogi and Yu in Ref.~\topcref.} If the top-pion is
lighter than the top quark, then
\eqn\tpibrate{\Gamma(t \ra \toppip b) \simeq
{(m_t^2 - M_{\pi_t}^2)^2 \over {32 \pi m_t F_t^2}} \ts.}
It is known that $B(t \ra W^+ b) = 0.87^{+0.13}_{-0.30}$ (stat.)
$^{+0.13}_{-0.11}$ (syst.)
\ref\twbrate{J.~Incandela, Proceedings of the 10th Topical Workshop on
Proton-Antiproton Collider Physics, Fermilab, edited R.~Raja and J.~Yoh,
p.~256 (1995).}.
At the $1\sigma$ level, then, $M_{\pi_t} \simge 150\,\gev$. At the
$2\sigma$ level, the lower bound is $100\,\gev$, but such a small branching
ratio for $t \ra W^+ b$ would require $\sigma(p\ol p \ra t \ol t)$ at the
Tevatron about~4 times the standard QCD value of
$4.75^{+0.63}_{-0.68}\,\pb$
\ref\topsig{S.~Catani, M.~Mangano, P.~Nason and L.~Trentadue,
CERN-TH/96-21, hep-ph/9602208 (1996).}.
The $t \ra \toppip b$ decay mode can be sought in high-luminosity runs
at the Tevatron and with moderate luminosity at the LHC. If $M_{\pi_t} <
m_t$, then $\toppip \ra c \ol b$ through $t$--$c$ mixing. It is also
possible, though unlikely, that $\tpip \ra t \ol s$ through $b$--$s$
mixing.

If $M_{\pi_t} > m_t$, then $\toppip \ra t \ol b$ and $\toppiz \ra \ol t t$
or $\ol c c$, depending on whether the top-pion is heavier or lighter than
$2m_t$. The main hope for discovering top-pions heavier than the top quark
seems to rest on the isotriplet of top-rho vector mesons, $\rho_t^{\pm,0}$.
It is hard to estimate $M_{\rho_t}$; it may lie near $2m_t$ or closer to
$\Lambda_t = \CO(1\,\tev)$. They are produced in hadron collisions just as
the corresponding color-singlet technirhos (Eq.~(3.1) of~I). The conventional
expectation is that they decay as $\rho_t^{\pm,0} \ra \toppipm \toppiz$,
$\toppip \toppim$. Then, the top-pion production rates may be estimated
from Eqs.~(3.2) and~(3.5) of~I with $\atro = 2.91$ and $\CC_{AB} =
1$. The rates are not large, but the distinctive decays of top-pions help
suppress standard model backgrounds.

Life may not be so simple, however. The $\rho_t$ are not completely
analogous to the $\rho$-mesons of QCD and technicolor because topcolor is
broken near $\Lambda_t$. Thus, for distance scales between $\Lambda^{-1}_t$
and $1\,\gev^{-1}$, top and bottom quarks do not experience a growing
confining force. Instead of $\rho_t \ra \toppi\toppi$, it is also possible
that $\rho_t^{\pm,0}$ fall apart into their constituents $t \ol b$, $b \ol
t$ and $t \ol t$. The $\rho_t$ resonance may be visible as a significant
increase in $t \ol b$ production, but it won't be in $t \ol t$.\foot{I
thank John Terning for inspiring this discussion of $\rho_t$ decays.}

The $V_8$ colorons of broken $SU(3)$ topcolor are readily produced in
hadron collisions. They are expected to have a mass of 1/2--1~TeV.
Colorons couple with strength $-g_S \cot\xi$ to quarks of the two light
generations and with strength $g_S \tan\xi$ to top and bottom quarks, where
$\tan\xi \gg 1$~\hp. Their decay rate is
\eqn\widveight{
\Gamma_{V_8} = {\aqcd \Mv \over {6}}\ts
\biggl\{4\cot^2\xi + \tan^2\xi \left(1 + \beta_t
(1-m_t^2/\Mv^2)\right)\biggr\} \ts.}
where $\beta_t = \sqrt{1 - 4 m_t^2/\Mv^2}$. Colorons may then appear as
resonances in $b \ol b$ and $t \ol t$ production. For example, the
$\CO(\aqcd)$ cross section for $\ol q q \ra \ol t t$ becomes
\eqn\qqbttb{
{d \hat \sigma(\ol q q\ra \ol t t) \over {d z}} =
{\pi \aqcd^2 \beta_t \over {9 \shat}} \ts \bigl(2 - \beta_t^2 +
\beta_t^2 z^2\bigr) \ts \biggl|1 - \ts{\shat \over
{\shat - \Mv^2 + i \sqrt{\shat} \ts \Gamma_{V_8}}}\biggr|^2 \ts.}
For completeness, the $gg \ra \ol t t$ rate is
\eqn\ggttb{\eqalign{
{d \hat \sigma(gg \ra \ol t t) \over {d z}} =
{\pi \aqcd^2 \beta_t \over {6 \shat}} \ts
&\biggl\{{1 + \beta_t^2 z^2 \over {1 - \beta_t^2 z^2}} -
{(1-\beta_t^2)^2 \ts (1 + \beta_t^2 z^2) \over{(1-\beta_t^2 z^2)^2}}
- {\tx{{9\over{16}}}} (1 + \beta_t^2 z^2) \cr
& + {1-\beta_t^2 \over{1-\beta_t^2 z^2}} \ts (1 -
{\tx{{1\over{8}}}} \beta_t^2 + {\tx{{9\over{8}}}} \beta_t^2 z^2)
\biggr\} \ts. \cr}}
A description of the search and preliminary mass limits for colorons and
other particles decaying to $\ol b b$ and $\ol t t$ are given in
Ref.~\ref\cdfcoloron{R.~M.~Harris, Proceedings of the 10th Topical Workshop
on Proton-Antiproton Collider Physics, Fermilab, edited R.~Raja and J.~Yoh,
p.~72(1995).}.

Colorons have little effect on the standard dijet production rate. The
situation may be very different for the $Z'$ boson of the broken strong
$U(1)$ interaction.\foot{This interaction differentiates between top and
bottom quarks, helping the former develop a large mass while keeping the
latter light.} In Ref.~\tctwoklee\ a scenario for topcolor was developed in
which it is natural that $Z'$ couples strongly to the fermions of the first
two generations as well as those of the third. The $Z'$ probably is heavier
than the colorons, roughly $\Mzp =1$--$3\,\tev$. Thus, at subprocess
energies well below $\Mzp$, the interaction of $Z'$ with all quarks is
described by a contact interaction, just what is expected for quarks with
substructure at a scale of a few~TeV. This leads to an excess of
jets at high $E_T$ and invariant mass
\ref\elp{E.~J.~Eichten, K.~Lane and M.~E.~Peskin, Phys.~Rev.~Lett.~{\bf
50}, 811 (1983)}, \ref\ehlq{E.~Eichten, I.~Hinchliffe, K.~Lane and C.~Quigg,
Rev.~Mod.~Phys.~{\bf 56}, 579 (1984).}.
An excess in the jet-$E_T$ spectrum consistent with $\Lambda = 1600\,\gev$
has been reported by the CDF Collaboration
\ref\cdfjets{F.~Abe, et al., The CDF Collaboration, Phys.~Rev~Lett~{\bf
77}, 438 (1996).}.
It remains to be seen whether it is due to topcolor or any other new
physics. As with quark substructure, the angular and rapidity distributions
of the high-$E_T$ jets induced by $Z'$ should be more central than
predicted by QCD. The $Z'$ may also produce an excess of high invariant
mass $\ell^+\ell^-$. It will be interesting to compare limits on contact
interactions in the Drell-Yan process with those obtained from jet
production.

The topcolor $Z'$ will be produced directly in $\ol q q$ annhilation in LHC
experiments. Because $Z'$ may be strongly coupled to so many fermions,
including technifermions in the LHC's energy range, it is likely to be very
broad. The development of TC2 models is at such an early stage that the
$Z'$ couplings, its width and branching fractions, cannot be predicted with
confidence. These studies are underway and we hope for progress on these
questions in the coming year.

\newsec{Signatures for Quark and Lepton Substructure}

The presence of three generations of quarks and leptons, apparently
identical except for mass, strongly suggests that they are composed of
still more fundamental fermions, often called ``preons''. It is clear that,
if preons exist, their strong interaction energy scale $\Lambda$ must be
much greater than the quark and lepton masses. Long ago, 't~Hooft figured
out how interactions at high energy could produce essentially massless
composite fermions: the answer lies in unbroken chiral symmetries of the
preons {\it and} confinement by their strong ``precolor'' interactions
\ref\thooft{G.~'t~Hooft, in {\it Recent Developments in
Gauge Theories}, edited by G.~'t~Hooft, et al. (Plenum, New York, 1980).}.
There followed a great deal of theoretical effort to construct a realistic
model of composite quarks and leptons (see, e.g.,
Ref.~\ref\comp{S.~Dimopoulos, S.~Raby and L.~Susskind, Nucl.~Phys.~{\bf
B173}, 208 (1980)\semi
M.~E.~Peskin, Proceedings of the 1981 Symposium on Lepton and Photon
Interactions at High Energy, edited by W.~Pfiel, p.~880 (Bonn, 1981) \semi
I.~Bars, Proceedings of the Rencontres de Moriond, {\it Quarks, Leptons and
Supersymmetry}, edited by Tranh Than Van, p.~541 (1982).}) which,
while leading to valuable insights on chiral gauge theories, fell far short
of its main goal.

In the midst of this activity, it was pointed out that the existence of
quark and lepton substructure will be signalled at energies well below
$\Lambda$ by the appearance of four-fermion ``contact'' interactions which
differ from those arising in the standard model
\ref\elp{E.~J.~Eichten, K.~Lane and M.~E.~Peskin, Phys.~Rev.~Lett.~{\bf
50}, 811 (1983).},
\ref\ehlq{E.~Eichten, I.~Hinchliffe, K.~Lane and C.~Quigg,
Rev.~Mod.~Phys.~{\bf 56}, 579 (1984).}.
These interactions are induced by the exchange of preon bound states and
precolor-gluons. The main constraint on their form is that they must be
$SU(3) \otimes SU(2) \otimes U(1)$ invariant because they are generated by
forces operating at or above the electroweak scale. These contact
interactions are suppressed by $1/\Lambda^2$, but the coupling parameter of
the exchanges---analogous to the pion-nucleon and rho-pion couplings---is
not small. Thus, the strength of these interactions is conventionally taken
to be $\pm 4\pi/\Lambda^2$. Compared to the standard model, contact
interaction amplitudes are then of relative order $\shat/\aqcd \Lambda^2$
or $\shat/\alpha_{EW} \Lambda^2$. The appearance of $1/\alpha$ and the
growth with $\shat$ make contact-interaction effects the lowest-energy
signal of quark and lepton substructure. They are sought in jet production
at hadron and lepton colliders, Drell-Yan production of high invariant mass
lepton pairs, Bhabha scattering, $e^+e^- \ra \mu^+\mu^-$ and $\tau^+\tau^-$
\ref\pdg{For current collider limits on substructure, see
Ref.~\cdfjets\ and the Review of Particle Physics, Particle Data Group,
{\it Phys.~Rev.}~{\bf D54}, 1, (1996).},
atomic parity violation
\ref\rosner{J.~Rosner, Phys.~Rev.~{\bf D53}, 2724 (1996), and references
therein.},
and polarized M{\o}ller scattering
\ref\kumar{K.~Kumar, E.~Hughes, R.~Holmes and P.~Souder, ``Precision Low
Energy Weak Neutral Current Experiments'', Princeton University (October
30, 1995), to appear in Modern Physics Letters~A.}.
Here, we concentrate on jet production and the Drell-Yan process at hadron
colliders.

The contact interaction most used so far to parameterize limits on the
substructure scale $\Lambda$ is the product of two left-handed electroweak
isoscalar quark and lepton currents. Collider experiments can probe values
of $\Lambda$ in the 2--5~TeV range (Tevatron) to the 15--20~TeV range (LHC;
see Refs.~\ehlq\ and \ref\gemtdr{GEM Technical Design Report, Chapter~2.6,
GEM~TN-93-262, SSCL-SR-1219; submitted to the Superconducting Super
Collider Laboratory, (April 30, 1993)\semi K.~Lane, F.~Paige, T.~Skwarnicki
and J.~Womersley, {\it Simulations of Supercollider Physics}, Boston
University preprint BUHEP-94-31, Brookhaven preprint BNL-61138,
hep-ph/9412280 (1994), to appear in Physics Reports.}). If $\Lambda$ is to
be this low, the contact interaction must be flavor-symmetric, at least for
quarks in the first two generations, to avoid large $\Delta S = 2$ and,
possibly, $\Delta B_d = 2$ neutral current interactions. We write it as
\eqn\Lcomp{
\CL^0_{LL} = {4 \pi \eta \over {2 \Lambda^2}}
\sum_{i,j=1}^3 \ts \left( \sum_{a=1}^3 \ol q_{aiL} \gamma^\mu q_{aiL} +
\CF_\ell \ts \ol \ell_{iL} \gamma^\mu \ell_{iL} \right)
\ts \left( \sum_{b=1}^3 \ol q_{bjL} \gamma_\mu q_{bjL} +
\CF_\ell \ts \ol \ell_{jL} \gamma_\mu \ell_{jL} \right) \ts.}
Here, $\eta=\pm 1$; $a,b = 1,2,3$ labels color; $i,j = 1,2,3$ labels the
generations, and the quark and lepton fields are isodoublets, $q_{ai} =
(u_{ai}, d_{ai})$ and $\ell_i = (\nu_i, e_i)$ The real factor $\CF_\ell$ is
inserted to allow for different quark and lepton couplings, but it is
expected to be $\CO(1)$. The factor of $\half$ in the overall strength of
the interaction avoids double-counting interactions and amplitudes.

The color-averaged jet subprocess cross sections, modified for the
interaction $\CL^0_{LL}$, are given in leading order in $\aqcd$ by (these
formulas correct errors in Ref.~\ehlq)
\eqn\sigcompjet{\eqalign{
& {d \hat\sigma(q_i q_i \ra q_i q_i) \over {dz}} =
{d \hat\sigma(\ol q_i \ol q_i \ra \ol q_i \ol q_i) \over {dz}} \cr
& = {\pi \over {2 \shat}}
\biggl\{ {4 \over {9}} \aqcd^2 \left[{\uhat^2 + \shat^2 \over {\that^2}} +
{\that^2 + \shat^2 \over {\uhat^2}} -
{2 \over {3}} {\shat^2 \over {\that \uhat}} \right]
+ {8 \over {9}} \aqcd {\eta \over {\Lambda^2}} \left[ {\shat^2 \over
{\that}} + {\shat^2 \over {\uhat}} \right] + {8 \over {3}} {\shat^2 \over
{\Lambda^4}} \biggr\} \ts; \cr\cr
& {d \hat\sigma(q_i \ol q_i \ra q_i \ol q_i) \over {dz}}\cr
& = {\pi \over {2 \shat}}
\biggl\{ {4 \over {9}} \aqcd^2 \left[{\uhat^2 + \shat^2 \over {\that^2}} +
{\uhat^2 + \that^2 \over {\shat^2}} -
{2 \over {3}} {\uhat^2 \over {\shat \that}} \right]
+ {8 \over {9}} \aqcd {\eta \over {\Lambda^2}} \left[ {\uhat^2 \over
{\that}} + {\uhat^2 \over {\shat}} \right] + {8 \over {3}} {\uhat^2 \over
{\Lambda^4}} \biggr\} \ts; \cr\cr
& {d \hat\sigma(q_i \ol q_i \ra q_j \ol q_j) \over {dz}} = 
{\pi \over {2 \shat}}
\biggl\{ {4 \over {9}} \aqcd^2 \left[{\uhat^2 + \that^2 \over {\shat^2}}
\right] + {\uhat^2 \over {\Lambda^4}} \biggr\} \ts; \cr\cr
& {d \hat\sigma(q_i \ol q_j \ra q_i \ol q_j) \over {dz}} =
{\pi \over {2 \shat}}
\biggl\{ {4 \over {9}} \aqcd^2 \left[{\uhat^2 + \shat^2 \over {\that^2}}
\right] + {\uhat^2 \over {\Lambda^4}} \biggr\} \ts; \cr\cr
& {d \hat\sigma(q_i q_j \ra q_i q_j) \over {dz}} =
{d \hat\sigma(\ol q_i \ol q_j \ra \ol q_i \ol q_j) \over {dz}} =
{\pi \over {2 \shat}}
\biggl\{ {4 \over {9}} \aqcd^2 \left[{\uhat^2 + \shat^2 \over {\that^2}}
\right] + {\shat^2 \over {\Lambda^4}} \biggr\}\ts .\cr}}
For this LL-isoscalar interaction, the interference term ($\eta/\Lambda^2$)
in the hadron cross section is small and the sign of $\eta$ is not very
important. Interference terms may be non-negligible in contact interactions
with different chiral, flavor, and color structures. In all cases, the main
effect of substructure is to increase the proportion of centrally-produced
jets. If this can be seen in the jet angular distribution, it will be
important for confirming the presence of contact interactions.\foot{This is
true regardless of the dynamical origin of the contact interaction.}

The modified cross sections for the Drell-Yan process $\ol q_i q_i \ra
\ell_j^+ \ell_j^-$ is
\eqn\sigcomplept{
{d\hat\sigma(\ol q_i q_i \ra \ell_j^+ \ell_j^-) \over {dz}} = {\pi
\alpha^2 \over {6 \shat}} \ts \left[ \CA_i(\shat) \left({\uhat \over
{\shat}} \right)^2 +
\CB_i(\shat) \left({\that \over {\shat}} \right)^2 \right] \ts ,}
where
\eqn\cacb{\eqalign{
\CA_i(\shat) = &\biggl[ Q_i + {4 \over {\sin^2 2 \thw}}
\left(T_{3i} - Q_i \sin^2 \thw \right) \left(\half - \sin^2 \thw
\right)
\left({\shat \over {\shat - M_Z^2}}\right) - {\CF_\ell \eta \shat \over
{\alpha \Lambda^2}} \biggr]^2 \cr
+&\biggl[ Q_i + Q_i \tan^2 \thw \left({\shat \over {\shat - M_Z^2}}\right)
\biggr]^2 \ts ; \cr
\CB_i(\shat) = &\biggl[ Q_i - {1 \over{\cos^2 \thw}} 
\left(T_{3i} - Q_i \sin^2 \thw \right) \left({\shat \over {\shat - M_Z^2}}
\right) \biggr]^2 \cr
+&\biggl[ Q_i - {1 \over {\cos^2 \thw}} Q_i  \left(\half - \sin^2
\thw \right) \left({\shat \over {\shat - M_Z^2}}\right) \biggr]^2 \ts.
\cr}}

The angular distribution of the $\ell^-$ relative to the incoming quark is
an important probe of the contact interaction's chiral structure. Measuring
this distribution is easy in a $\ol p p$ collider such as the Tevatron
since the hard quark almost always follows the proton direction. If the
scale $\Lambda$ is high so that parton collisions revealing the contact
interaction are hard, the quark direction can also be determined with
reasonable confidence in a $pp$ collider. At the LHC, the quark in a $\ol q
q$ collision with $\sqrt{\shat/s} \simge 1/20$ is harder than the
antiquark, and its direction is given by the boost rapidity of the dilepton
system, at least 75\% of the time. The charges of $\CO(1\,\tev)$ muons can
be well-measured even at very high luminosity in the detectors being
designed for the LHC. These two ingredients are needed to insure a good
determination of the angular distribution~\gemtdr.

It is important to study the effects of contact interactions with chiral,
flavor and color structures different from the one in Eq.~\Lcomp. Such
interactions can give rise to larger (or smaller) cross sections for the
same $\Lambda$ because they have more terms or because they interfere more
efficiently with the standard model. Thus, it will be possible to probe
even higher values of $\Lambda$ for other structures. Other forms can also
give rise to $\ell^\pm\nu$ final states. Searching for contact interactions
in these modes is more challenging than in $\ell^+\ell^-$, but it is very
useful for untangling flavor and chiral structures~\gemtdr. Events are
selected which contain a single high-$p_T$ charged lepton, large missing
$E_T$, and little jet activity. Even though the parton c.m.~frame cannot be
found in this case, it is still possible to obtain information on the
chiral nature of the contact interaction by comparing the $\vert
\eta_{\ell^+} \vert$ and $\vert \eta_{\ell^-} \vert$ rapidity distributions
of the high-$p_T$ leptons. For example, if the angular distribution in the
process $d \ol u \ra \ell^- \ol \nu$ between the incoming $d$-quark and the
outgoing $\ell^-$ is $(1 + \cos\theta)^2$, then $\vert\eta_{\ell^-}\vert$
is pushed to larger values because the $d$-quark is harder than the $\ol
u$-quark and the $\ell^-$ tends to be produced forward. Correspondingly, in
$u \ol d \ra \nu \ell^+$, the $\vert\eta_{\ell^+}\vert$ distribution would
be squeezed to smaller values.

\newsec{Conclusions and Acknowledgements}

Many theorists are convinced that low-energy supersymmetry is intimately
connected with electroweak symmetry breaking and that its discovery is just
around the corner
\ref\susylett{see, e.g., P.~Ramond, et al., Letter to the Director of
Fermilab and the Co-Spokespersons of the CDF and D\O\ Collaborations,
(1994).}.
One often hears that searches for other sorts of TeV-scale physics are,
therefore, a waste of time. Experimentalists know better. The vast body of
{\it experimental} evidence favors no particular extension of the standard
model. Therefore, all plausible approaches must be considered. Detectors
must have the capability---and experimenters must be prepared---to discover
whatever physics is responsible for electroweak and flavor symmetry
breaking. To this end, we have summarized the principal signatures for
technicolor, extended technicolor and quark-lepton substructure. Table~1
lists sample masses for new particles and their production rates at the
Tevatron and LHC. We hope that this summary is useful to future in-depth
studies of strong TeV-scale dynamics.

\medskip

We are especially grateful to John Womersley and Robert Harris for
encouragement, advice and thoughtful readings of the manuscript. We are
indebted to those members of CDF and D\O\ who discussed their work with us
and otherwise helped us prepare our review: Tom Baumann, John Huth, Kaori
Maeshima, Wyatt Merritt and Jorge Troconiz. Finally, we thank Dimitris
Kominis for discussions on topcolor-assisted technicolor and for catching
several errors.

\vfil\eject

\listrefs

\centerline{\vbox{\offinterlineskip
\hrule\hrule
\halign{&\vrule#&
  \strut\quad#\hfil\quad\cr\cr
height4pt&\omit&&\omit&&\omit&&\omit&\cr\cr
&\hfill Process \hfill&&\hfill
Sample Mass (GeV)  \hfill&&\hfill $\sigma_{\rm TeV}$(pb) \hfill&&\hfill
$\sigma_{\rm LHC}$(pb) \hfill &\cr\cr
height4pt&\omit&&\omit&&\omit&&\omit&\cr\cr
\noalign{\hrule\hrule}
height4pt&\omit&&\omit&&\omit&&\omit&&\omit&&\omit&\cr\cr
&$\tro\ra W_L\tpi${$\ts ^{1}$}
&&\hfill $220(\tro)$, $100(\tpi)$\hfill&&\hfill$5$\hfill&&
\hfill$35$ \hfill& \cr\cr
\noalign{\hrule}
height4pt&\omit&&\omit&&\omit&&\omit&\cr\cr
&$\tro\ra\tpi\tpi${$\ts ^{1}$} &&\hfill
$220(\tro)$, $100(\tpi)$\hfill&&\hfill$5$\hfill&&
\hfill$25$\hfill&\cr\cr
\noalign{\hrule}
height4pt&\omit&&\omit&&\omit&&\omit&\cr\cr
&$gg \ra \tpiz \ra b \ol b${$\ts ^{2}$}&&\hfill
$100$\hfill&& \hfill$300/5000$\hfill&&\hfill
$7000/10^{5}$\hfill& \cr\cr
\noalign{\hrule}
height4pt&\omit&&\omit&&\omit&&\omit&\cr\cr
&$gg \ra \eta_T \ra t \ol t${$\ts ^{3}$}&&\hfill $400$\hfill&&
\hfill$3/3$\hfill&&\hfill
$2000/600$\hfill& \cr\cr
\noalign{\hrule}
height4pt&\omit&&\omit&&\omit&&\omit&\cr\cr
&$gg\ra\tpi\tpi${$\ts ^{4}$}&&\hfill $100$\hfill&&
\hfill$0.2$\hfill&&\hfill $600$\hfill& \cr\cr
\noalign{\hrule}
height4pt&\omit&&\omit&&\omit&&\omit&\cr\cr
&$\troct\ra \jet\ts\ts\jet${$\ts ^{5}$}&&\hfill $250(\troct)$\hfill&&
\hfill$700/5000$\hfill&&\hfill $1.5\times 10^{4}/1.5\times
10^{5}$\hfill& \cr\cr
%
%
height4pt&\omit&&\omit&&\omit&&\omit&\cr\cr
& &&\hfill $500(\troct)$\hfill&&
\hfill$10/40$\hfill&&\hfill $2000/6000$\hfill& \cr\cr
\noalign{\hrule}
height4pt&\omit&&\omit&&\omit&&\omit&\cr\cr
&$\troct\ra\pi_{T8}\pi_{T8}${$\ts ^{6}$}&&\hfill $550(\troct),
250(\octpi)$\hfill&& \hfill$2$\hfill&&\hfill $2000$\hfill& \cr\cr
\noalign{\hrule}
height4pt&\omit&&\omit&&\omit&&\omit&\cr\cr
&$\troct\ra \tpiql\tpilq${$\ts ^{6}$}&&\hfill$550(\troct),
200(\tpiql)$\hfill&& \hfill$2$\hfill&&\hfill $1000$\hfill& \cr\cr
\noalign{\hrule}
height4pt&\omit&&\omit&&\omit&&\omit&\cr\cr
&$V_8 \ra t \ol t${$\ts ^{7}$}&&\hfill $500$\hfill&&
\hfill$8/3$\hfill&&\hfill $100/600$\hfill& \cr\cr
\noalign{\hrule}
height4pt&\omit&&\omit&&\omit&&\omit&\cr\cr
&$\Lambda$ reach{$\ts ^{8}$}&&\hfill 5~TeV (TeV),
20~TeV (LHC)\hfill&& \hfill$10\,\fb^{-1}$\hfill&&\hfill
$100\,\fb^{-1}$\hfill& \cr\cr 
height4pt&\omit&&\omit&&\omit&&\omit&\cr\cr}
\hrule\hrule}}

\bigskip

\noindent Table 1. Sample cross sections for technicolor signatures at the
Tevatron and LHC. \hfil\break
Cross sections may vary by a factor of~10 for other masses and choices of
the parameters. $K$-factors of 1.5--2 are expected, but not included.
Signal over background rates are quoted as $S/B$. $\Ntc = 4$ in all
calculations; cross sections generally grow with $\Ntc$. 

\medskip

\noindent $^{1}$ $F_T = F_\pi/3 = 82\,\gev$ was used.

\noindent $^{2}$ $F_T = 50\,\gev$ used. Cross
section is integrated over $\CM_{b \ol b} = 90$--$110\,\gev$.

\noindent $^{3}$ $F_T = 50\,\gev$ and $m_t = 175\,\gev$ were used. The
greatly increased LHC cross section is due to the rapid growth of gluons at
small-$x$.

\noindent $^{4}$ Cross sections for a multiscale model with 250~GeV
$\octpi$ and 200~GeV $\tpiql$ intermediate states.

\noindent $^{5}$ Jet energy resolution of $\sigma(E)/E = 100\%/\sqrt{E}$
is assumed and cross sections integrated over $\pm \Gamma$ about resonance
peak. Jet angles are limited by $\cos \theta^* < \twothirds$ and $\vert
\eta_j \vert < 2.0$ (Tevatron) or 1.0 (LHC).

\noindent $^{6}$ Cross sections per channel are quoted.

\noindent $^{7}$ $\tan\xi = \sqrt{2\pi/3\aqcd}$ was used, corresponding to a
critical topcolor coupling strength.

\noindent $^{8}$ Estimated $\Lambda$ reaches in dijet and dilepton production
are for the indicated luminosities.

\bye